\documentclass[aps,prl,twocolumn,superscriptaddress]{revtex4-1}
\pdfoutput=1
\usepackage{graphicx}
\usepackage{color}
\usepackage{hyperref}
\usepackage{amsmath}
\usepackage{url}

\newcommand{\pasj}{Publ.~Astron.~Soc.~Japan}

\newcommand{\psl}{P^{\rm L}}

\newcommand{\bk}{{\bf k}}

\newcommand{\bptnw}{{B^{\rm PT}_{\rm nw}}}

\begin{document}

\title{Bispectrum as baryon acoustic oscillation interferometer}
\author{Hillary L. Child}
\email{childh@uchicago.edu}
\affiliation{HEP Division, Argonne National Laboratory, Lemont, Illinois 60439, USA}
\affiliation{Department of Physics, University of Chicago, Chicago, Illinois 60637, USA}
\author{Masahiro Takada}
\affiliation{Kavli Institute for the Physics and Mathematics of the
Universe (WPI), UTIAS, The University of Tokyo, Chiba 277-8583, Japan}
\author{Takahiro Nishimichi}
\affiliation{Kavli Institute for the Physics and Mathematics of the
Universe (WPI), UTIAS, The University of Tokyo, Chiba 277-8583, Japan}
\author{Tomomi Sunayama}
\affiliation{Kavli Institute for the Physics and Mathematics of the
Universe (WPI), UTIAS, The University of Tokyo, Chiba 277-8583, Japan}
\author{Zachary Slepian}
\affiliation{Einstein Fellow, Lawrence Berkeley National Laboratory, 1 Cyclotron Road, Berkeley, California 94720, USA}
\affiliation{Berkeley Center for Cosmological Physics, University of California, Berkeley, California 94720, USA}
\author{Salman Habib}
\affiliation{HEP Division, Argonne National Laboratory, Lemont, Illinois 60439, USA}
\affiliation{MCS Division, Argonne National Laboratory, Lemont, Illinois 60439, USA}
\author{Katrin Heitmann}
\affiliation{HEP Division, Argonne National Laboratory, Lemont, Illinois 60439, USA}
\affiliation{MCS Division, Argonne National Laboratory, Lemont, Illinois 60439, USA}

\date{\today}

\begin{abstract}
The galaxy bispectrum, measuring excess clustering of galaxy triplets, offers a probe of dark energy via baryon acoustic oscillations (BAOs). However up to now it has been severely underused due to the combinatorically explosive number of triangles. Here we exploit interference in the bispectrum to identify triangles that amplify BAOs. This approach reduces the computational cost of estimating covariance matrices, offers an improvement in BAO constraints equivalent to lengthening BOSS by 30\% and simplifies adding bispectrum BAO information to future large-scale redshift survey analyses.
 \end{abstract}

\maketitle

\section{Introduction}
The baryon acoustic oscillation (BAO) method exploits the imprint of sound waves in the prerecombination plasma on the late-time clustering of galaxies to measure the expansion history of the Universe and constrain the dark energy equation of state \cite{1998ApJ...504L..57E, 2003ApJ...594..665B, 2003PhRvD..68f3004H, 2003PhRvL..90i1301L, 2003ApJ...598..720S}. Applied to Baryon Oscillation Spectroscopic Survey (BOSS) data, the BAO method has offered 1\% distance constraints \cite{Ross:2016gvb, 2017MNRAS.470.2617A}; future surveys such as DESI \footnote{\url{http://desi.lbl.gov}} and Subaru PFS \cite{Takadaetal:14} promise to tighten these to subpercent precision. The BAO precision from the power spectrum $P(k)$ and two-point correlation function is further improved by reconstruction \cite{Eisensteinetal:07, 2009PhRvD..80l3501N, 2009PhRvD..79f3523P, Padmanabhanetal:12, Schmittfull:2015mja}, which uses the density field as sampled by galaxies to partially reverse the smearing effects of nonlinear structure formation on the BAO peak. 

Direct measurements of higher-point functions may yet provide an additional improvement on BAO constraints, and recent algorithms \cite{2015PhRvD..92h3532S, Slepian:2015qwa,Slepian:2015qza, Slepian:2017lpm,Sugiyama:2018yzo} have reduced the computational complexity of calculating three-point statistics. The BAO feature has recently been detected in the three-point correlation function \cite{Gaztanaga:2008sq, Slepian:2015hca, Slepian:2016kfz} and bispectrum $B(k_1, k_2, k_3)$ \cite{Pearson:2017wtw}. Like reconstruction, three-point information can improve constraints on the BAO scale: for example, Ref.~\cite{Slepian:2015qza} finds a 6\% improvement \cite{Slepian:2016kfz} using CMASS data. However the high number of triangles necessitates a large number of mock catalogs to directly compute covariance matrices. One approach to this challenge is an analytic covariance template \cite{Slepian:2015qza}, which improves the covariance matrix calculated from a smaller number of mocks. Alternatively, using only a diagonal covariance matrix \cite{2015MNRAS.451..539G} or measuring bispectra on only a subset of all possible triangles dramatically reduces the size of the covariance matrix. However, simple rules for selecting triangles (e.g. isosceles, or one side an integer multiple of another \cite{2011ApJ...739...85M,2015MNRAS.451..539G,2017MNRAS.465.1757G}) may be far from optimal for probing BAOs. 

Here we identify triangle configurations that maximize or minimize the BAO signal in the bispectrum, enabling precise BAO constraints with relatively few bispectrum measurements. Our approach allows intuitive visualization of bispectra as functions of a single variable, because we set two of the three wave numbers to depend on the first.

{\sl Perturbation Theory Model.--} We first explore the perturbation theory (PT) bispectrum $B^{\rm PT}$ to study the BAO feature's dependence on triangle configuration. In contrast to the power spectrum, the bispectrum depends on a closed triangle formed by the three wave vectors $(\bk_1,\bk_2,\bk_3)$. We consider the isotropic bispectrum, where six degrees of freedom are redundant, so we can specify a triangle by e.g. three wave numbers $(k_1, k_2, k_3)$. The tree-level matter bispectrum in real space (i.e., without redshift-space distortions) \citep{Scoccimarro:1997st} is 
\begin{equation}
\label{eqn:B0}
B^{\rm PT}(k_1, k_2, k_3) = 2 \psl(k_1)\psl(k_2)F_2(\bk_1, \bk_2) + \mbox{cyc.}, 
\end{equation}
with
\begin{multline}
\label{eqn:F2}
F_2(\bk_i, \bk_j) \\
=\frac{5}{7}+\frac{1}{2}\left(\frac{k_i}{k_j}+\frac{k_j}{k_i}\right)(\hat{\bk}_i\cdot\hat{\bk}_j)+\frac{2}{7}(\hat{\bk}_i\cdot\hat{\bk}_j)^2,
\end{multline}
where $\psl(k)$ is the {\em linear} matter power spectrum. We refer to $2\psl(k_1)\psl(k_2)F_2(\bk_1, \bk_2)$ as the precyclic term and to the terms denoted by cyc as the postcyclic terms. 

The linear power spectrum involves the square of the matter transfer function: 
\begin{equation}
\psl(k) = P_{\rm pri}(k) T_{\rm m}^2(k)\\
\end{equation}
where $P_{\rm pri}(k)$ is the primordial power spectrum.

We split the transfer function into smooth and oscillatory pieces as $T_{\rm m}(k)= T_{\rm sm}(k)+\epsilon(k)j_0(k\tilde{s})$, where $T_{\rm sm}(k)$ and $\epsilon(k)$ are smooth functions of $k$ \cite{EisensteinHu:98}, the oscillations come from BAOs, and $\epsilon$ is small (as $\Omega_{\rm b}/\Omega_{\rm m} \ll 1$). 
The effective sound horizon $\tilde{s}(k)$ vanishes at low $k$ and is within 1\% of the true sound horizon for $k \gtrsim 0.1\,h/{\rm Mpc}$ \cite{EisensteinHu:98}; we use $\tilde{s}_f \equiv \tilde{s}(k_f)$ at the fiducial wave number $k_f=0.2\,h/{\rm Mpc}$. The spherical Bessel function $j_0(k\tilde{s}_f)$ has wavelength $2\pi/\tilde{s}_f$ with a decaying envelope $1/(k\tilde{s}_f)$. In each term of $B^{\rm PT}$ [Eq.~\eqref{eqn:B0}], 
BAOs enter through the product of transfer functions $T_{\rm m}^2(k_i)T_{\rm m}^2(k_j)$. To leading order in $\epsilon$, the oscillatory part scales as
\begin{multline}
\hspace{-0.5em}\frac{\left[T_{\rm m}^2(k_i)T_{\rm m}^2(k_j)\right]_{\rm osc}}
{T_{\rm sm}^2(k_i)T_{\rm sm}^2(k_j)} \\
\propto\frac{\epsilon(k_i)}{T_{\rm sm}(k_i)}\frac{\sin(k_i\tilde{s}_f)}{k_i\tilde{s}_f}+\frac{\epsilon(k_j)}{T_{\rm sm}(k_j)}\frac{\sin(k_j\tilde{s}_f)}{k_j\tilde{s}_f}. 
\end{multline}
In the precyclic term, defined below Eq.~\eqref{eqn:B0}, interference will therefore depend on the phase difference between $k_1$ and $k_2$ in units of $2\pi/\tilde{s}_f$ (i.e. the BAO wavelength in Fourier space), motivating the triangle parametrization 
\begin{equation}
k_1,\hspace{1em} k_2 - k_1 = \delta \left(\frac{\pi}{\tilde{s}_f}\right), \hspace{1em} \cos\theta=\hat{\bk}_1\cdot\hat{\bk}_2.
\end{equation}
When $\delta$ is an odd integer, interference in the precyclic term is perfectly destructive, suppressing BAOs; for even integer $\delta$, interference is perfectly constructive, amplifying BAOs.
At fixed $\delta$, $\theta$ determines both the phase structure of the postcyclic terms and, through $F_2$, the relative contributions of all three terms to $B^{\rm PT}$. We compute $B^{\rm PT}$ for a range of $k_2(k_1)$ and all possible triangular $k_3(k_1, k_2)$ and select the configurations that maximize or minimize the BAO features. We refer to these as ``constructive'' and ``destructive'' configurations, where ``configuration" means a set of triangles with fixed $\delta$ and $\theta$ over a range of $k_1$.

\begin{figure}
	\centerline{
    \includegraphics{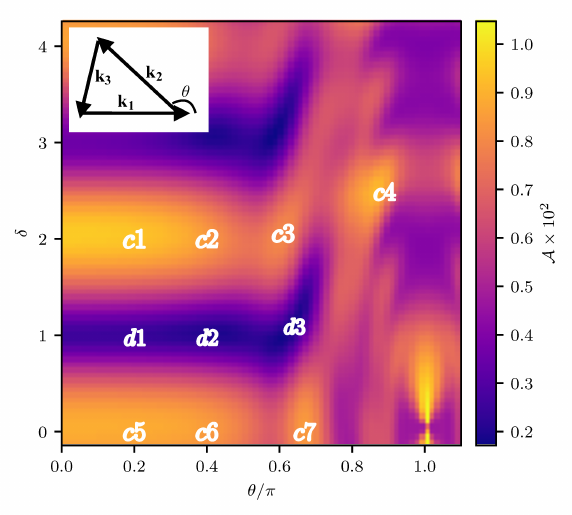}}
	\caption{The standard deviation $\mathcal{A}$ of 
    the bispectrum BAO feature in triangle configurations parametrized by $(\delta, \theta)$; the
     inset shows the definition of $\theta$. 
    We measure labeled configurations in simulations to improve constraints on the BAO scale.
\label{fig:delta_theta_rms}}
\end{figure}
The amplitude of the bispectrum does include BAOs and redshift-space distortion information, but here we focus on the oscillatory behavior, isolated in the ratio 
\begin{equation}
\label{eqn:R}
R(k_1,\delta,\theta) = \frac{B^{\rm PT}(k_1,\delta,\theta)}{\bptnw(k_1,\delta,\theta)}.
\end{equation}
$B^{\rm PT}$ involves a power spectrum from CAMB~\citep{Lewis:1999bs}; we use a flat $\Lambda$CDM cosmology with ${\rm \Omega}_{\rm m} = 0.2648$, ${\rm \Omega}_{\rm b}h^2 = 0.02258$, $n_s = 0.963$, $\sigma_8 = 0.8$, and $h=0.71$.  In this cosmology, $\tilde{s}_f \approx 109.5\,{\rm Mpc}/h$ via the fit of \cite{EisensteinHu:98}. $\bptnw$ is the bispectrum computed using the no-wiggle power spectrum \cite{EisensteinHu:98}.

Constructive interference enhances ``wiggles'' and increases the standard deviation, denoted $\mathcal{A}$, of Eq.~(\ref{eqn:R}). We thus quantify BAO interference by
\begin{equation}
\mathcal{A}^2(\delta,\theta) \equiv \int_{0.01}^{0.2} \left[R(k_1,\delta,\theta) - \bar{R}(\delta,\theta) \right]^2 
\frac{\mathrm{d}k_1}{[h/{\rm Mpc}]},
\end{equation}
where $\bar{R}(\delta, \theta)$ is the mean of $R(k_1, \delta, \theta)$ on the same range, $0.01 \leq k_1/[h/{\rm Mpc}] \leq 0.2$. Beyond the lower edge of this range, cosmic variance will limit the usefulness of bispectrum measurements, while beyond the upper end, Silk damping as well as late-time nonlinear structure formation damp BAOs. We display $\mathcal{A}$ in Fig.~\ref{fig:delta_theta_rms}. This figure is a guide to identify constructive bispectrum configurations, where we expect the strongest BAO signal. 

\begin{figure*}
	\centerline{
    \includegraphics{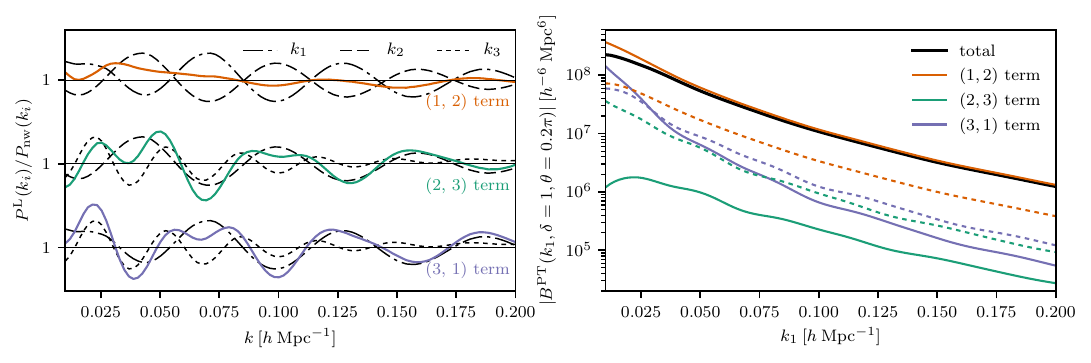}}
	\caption{For a destructive configuration such as $d_1$ (with $\delta=1$ and $\theta=0.2\pi$), BAOs are out of phase in the dominant term or terms of $B^{\rm PT}$.
    Left: Interference of the BAO feature for each pair of wave numbers. Black curves show the ratio of the linear to the no-wiggle power spectrum, $\psl(k_i)/P_{\rm nw}(k_i)$, for a single wave number; the product of each pair of ratios is shown in color. For example, the oscillations in the $k_1$ and $k_2$ terms are out of phase, so the product $\psl(k_1)\psl(k_2)/[P_{\rm nw}(k_1)P_{\rm nw}(k_2)]$ (orange) shows nearly no oscillation. 
    Right: $B^{\rm PT}$ and its three terms: the products of power spectra weighted by the $F_2$ kernel (solid, color curves) that sum to the total $B^{\rm PT}$ (solid, black curves). Dashed curves show products of power spectra before weighting by the $F_2$ kernel. The $(1,2)$ term (orange) is the primary contribution to $B^{\rm PT}$; because the BAOs destructively interfere in this term, $B^{\rm PT}$ shows little oscillation and so the configuration is destructive.}
\label{fig:wiggles_contributions}
\end{figure*}

Three effects combine to determine the oscillations' amplitude in each configuration. First, the broadband behavior of $\psl(k)$ sets the magnitude of each $\psl(k_i)\psl(k_j)$ permutation. Second, the multiplication by $F_2(k_i, k_j)$ modulates this overall magnitude. Thus in any configuration the dominant term will be the one with large $\psl(k_i)\psl(k_j)$, further enhanced by a large $F_2(k_i, k_j)$. Third, the amplitude of oscillation in each $\psl(k_i)\psl(k_j)$ depends on whether the oscillations in $\psl(k_j)$ are in or out of phase with the oscillations in $\psl(k_i)$. Therefore, the phase shift in the dominant term (or terms) sets whether the configuration is constructive (amplifying BAOs) or destructive (suppressing BAOs). As an example, the contributions to $B^{\rm PT}$ in a destructive configuration, $d_1$, are shown in Fig.~\ref{fig:wiggles_contributions}. 

{\sl Simulations.--} Motivated by the above perturbative analysis, we now explore the power of our interferometric approach to constrain the BAO scale in full $N$-body simulations, which accurately solve for the nonlinear structure formation giving rise to the bispectrum.  We measure halo power spectra and bispectra from four $z=0.55$ MockBOSS halo catalogs \cite{Sunayamaa:2015aba} with box size $L = 4000\,{\rm Mpc}/h$ and particle mass $m_p = 6.8 \times 10^{10}\,{\rm M}_\odot/h$  using a friends-of-friends (FOF) finder with linking length $0.168$. Our threshold halo mass is $M_{\rm h}\ge 10^{13}\,{\rm M}_\odot/h$ (100 particles in the FOF group), giving number density $n = 3.8 \times 10^{-4} h^3/ {\rm Mpc}^{3}$. Our halo sample is thus roughly matched to the number density ($3 \times 10^{-4} h^3/ {\rm Mpc}^{3}$) and average redshift ($z=0.57$) of the BOSS CMASS galaxy sample \cite{2013AJ....145...10D}. To measure $B(k_1,\delta,\theta)$ and $P(k)$ from the halo catalog, we use an FFT-based algorithm \citep{2015JCAP...05..007B} with aliasing correction via interlacing \citep{2016MNRAS.460.3624S}.

Constraining power differs between constructive and destructive configurations, so we focus on ten triangle configurations shown in Fig.~\ref{fig:delta_theta_rms}: seven constructive ($c_1,\dots,c_7$) and three destructive ($d_1,d_2,d_3$). We ignore the maximum-amplitude configurations with $\theta=\pi$; these correspond to three collinear points in Fourier space, suggesting they are less independent from the power spectrum \cite{2017JCAP...06..022C}. Additionally, the small values of $k_3 \sim 10^{-3}\,h/{\rm Mpc}$ for these configurations make them subject to cosmic variance in practice, as there are few modes this large in typical surveys (e.g. BOSS or DESI).

Each configuration comprises 19 triangles because $k_1$ varies in 19 bins between $0.01\,h/{\rm Mpc}$ and $0.2\,h/{\rm Mpc}$.  In each bin, the other two wave numbers $k_2$ and $k_3$ are computed from $k_1$ according to the parameters $\delta$ and $\theta$, held fixed for any one configuration. Thus across the ten configurations we measure 190 triangles in total. This number represents roughly one-fourth of the 805 triangles that can be formed from three sides in our range and binning. We note that some triangles we measure have one or two sides exceeding $0.2 \, h/{\rm Mpc}$; the maximum $k_2$ we measure is $0.3 \, h/{\rm Mpc}$, while $k_3$ reaches $0.45 \, h/{\rm Mpc}$. Little BAO information comes from these scales, as the BAO feature is heavily damped for $k \gtrsim 0.2\, h/{\rm Mpc}$.

\begin{figure*}
	\centerline{
    \includegraphics{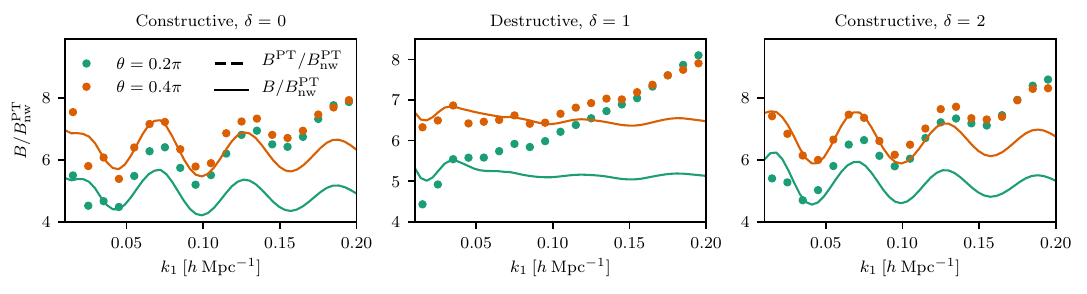}}
	\caption{Bispectrum results for four constructive (left and right panels) and two destructive (middle panel) configurations measured from the MockBOSS simulations. Points show the ratio of the simulated bispectrum $B$ to the PT no-wiggle bispectrum $\bptnw$.
    Curves show the ratio of the PT bispectrum $B^{\rm PT}$ to the no-wiggle $\bptnw$. Each PT curve $B^{\rm PT}/\bptnw$ oscillates about $B/\bptnw = 1$ but has been multiplied by a constant to account for linear bias and allow comparison with corresponding measurements.
   }
\label{fig:sim_destructive_constructive}
\end{figure*}

Figure~\ref{fig:sim_destructive_constructive} shows that simulated bispectra for destructive configurations have no BAO feature, while  pronounced oscillations appear in the constructive configuration bispectra. Nonlinear effects are more important for the configurations with larger $\theta$: as $\theta$ rises, so does $k_3$. As $\theta$ increases, we therefore expect the measured $B(k_1, \delta, \theta)$ to depart from $B^{\rm PT}(k_1, \delta, \theta)$ at smaller $k_1$,  reflecting the increasingly nonlinear behavior of $P(k_3)$. 

We perform a Fisher matrix analysis to estimate the improvement in BAO precision offered by our bispectrum approach. We use the 1$\sigma$ uncertainty $\sigma_\alpha$ in the shift parameter $\alpha$ as a measure of constraining power. We introduce nuisance parameters $A_P$ and $\mathbf{A_B}$ to model possible uncertainties (e.g. due to halo bias) in the amplitude of $P(k)/P_{\rm nw}(k)$ and each $B(k_1, \delta, \theta)/\bptnw(k_1, \delta, \theta)$, respectively. 
That is, our models for the power spectrum and the bispectrum for the $i$th triangle configuration are
\begin{align}
&P(k)/P_{\rm nw}(k) = A_P P(\alpha k)/P_{\rm nw}(\alpha k),\\
&R_{i}(k_1, \delta, \theta) = A_{B, i} R_i(\alpha k_1, \delta, \theta),
\end{align}
where $R_i$ denotes the ratio of the bispectrum $B_i$ to the corresponding no-wiggle bispectrum as in Eq.~\eqref{eqn:R}.
We test an additional set of nuisance parameters such that $P(k)/P_{\rm nw}(k)= A_P \left[P(\alpha k)/P_{\rm nw}(\alpha k)\right] +C_p k$ and $R_i(k_1, \delta, \theta) = A_{B, i} R_i(\alpha k_1, \delta, \theta) + C_{B, i}k_1$. The parameters $C_P$ and $\mathbf{C_B}$ are motivated by the behavior of the $R_i$ shown in Fig.~\ref{fig:sim_destructive_constructive}, which rise linearly as $k_1$ increases. Marginalizing over all four nuisance parameters changes our results by $\lesssim 10\%$, similar to other sources of error in our rough Fisher analysis.

The Fisher matrix is 
\begin{equation}
F_{\mu\nu} = \frac{\partial {\bf d}^T}{\partial p_\mu} {\bf C}^{-1}
 \frac{\partial {\bf d}}{\partial p_{\nu}},
\end{equation}
where $p_\mu\in \{\alpha, A_P, \mathbf{A}_B\}$, 
${\bf d}$ is the data vector of power spectrum and bispectrum measurements divided by their no-wiggle analogs, and ${\bf C}^{-1}$ is the 
inverse covariance matrix. 
The dimension of the data vector is 209: 19 bins for $P(k)$ plus 190 triangles. To estimate the covariance matrix, we subdivide each of the four $(4\,{\rm Gpc}/h)^3$ simulations into subvolumes 
of $(500\,{\rm Mpc}/h)^3$. The resulting 2048 subvolumes exceed the 209 power spectrum and bispectrum bins by roughly an order of magnitude, and the derived covariance matrix is well conditioned.
When computing the inverse of the covariance matrix, 
we include the correction factor in 
\cite{Hartlap:2006kj} to obtain its unbiased 
estimate.  

To compute the partial derivatives of $P(k)/P_{\rm nw}(k)$ and $B(k_1)/B^{\rm PT}_{\rm nw}(k_1)$ with respect to $\alpha$, we stretch and compress the simulation box by a factor $(1\pm \epsilon)$ with $\epsilon=0.05$, recompute the measured $P(k)$ and $B(k_1, \delta, \theta)$, and correct for the change in amplitude due to the change in box volume. This correction, the Jacobian of the integration measure between Fourier and real space, is $(1+\epsilon)^3$ for $P(k)$ and $(1+\epsilon)^6$ for the bispectrum \cite{Seo:2003pu,2014MNRAS.444.1400N}. 
With these amplitude shifts corrected, the power spectrum oscillations are simply stretched and compressed relative to the standard box. The oscillations as a function of $k_1$ are stretched in the bispectrum measurements as well. But even configurations where oscillations are suppressed contain BAO scale information: when $\alpha \neq 1$, configurations are selected based on an incorrect estimate of the sound horizon, so the measured configurations depart from the desired configurations. Consequently, the \emph{amplitude} of oscillation in $B(k_1)/\bptnw(k_1)$ changes as well as the frequency. 
 This behavior provides a small additional constraint on $\alpha$---even destructive configurations contain BAO scale information. 

The variance in $\alpha$ including marginalization over the amplitude parameters
is 
\begin{equation}
\sigma_\alpha^2 = ({\bf F}^{-1})_{\alpha \alpha}.
\end{equation}
We compute $\sigma_\alpha$ for $P(k)/P_{\rm nw}(k)$ alone and for $P(k)/P_{\rm nw}(k)$ combined with $B(k_1)/\bptnw(k_1)$ for different sets of triangle configurations. We also compute $P(k)/P_{\rm nw}(k)$ for the postreconstruction $P_{\rm r}(k)$, where reconstruction uses the algorithm of \cite{Padmanabhanetal:12} with bias measured from $P(k)$ and a smoothing scale of $R_s = 15\, {\rm Mpc}/h$. Our results are unchanged at the percent level with a smaller smoothing scale of $R_s = 5\, {\rm Mpc}/h$. 
Both reconstruction and the addition of bispectrum information decrease the uncertainty in $\alpha$. Computing the improvement in $\sigma_\alpha$ relative to the prereconstruction, $P(k)$-only constraint, we find that the relative improvement depends on the number and choice of triangle configurations. Three destructive configurations (57 triangles) only improve $\sigma_\alpha$ by less than 3\%, but a different 57 triangles in three constructive triangle configurations ($c_1$, $c_2$, and $c_3$) improve $\sigma_\alpha$ by roughly 8\%. With an additional four constructive configurations (for a total of 133 triangles in seven constructive configurations), the improvement reaches 12\%. The marginal improvement from each configuration depends as well on its covariance with previously selected configurations; future work will explore this dependence more fully. All ten triangle configurations (seven constructive and three destructive) provide an improvement of roughly 14\% over the precision from $P(k)$ alone. This initial choice of ten configurations does not exhaust the information available in the bispectrum; the seventh and last constructive configuration, for example, still raises the improvement in constraints by about 20\%. We thus anticipate that opportunities remain to further improve constraints on $\alpha$ by strategically selecting triangle configurations with minimal covariance. We will study the covariance structure of the bispectrum and its implications for triangle selection in detail in future work.

Fisher analysis also gives an estimate of how constraints from the bispectrum compare to reconstruction, currently the best available technique to improve BAO precision. Following the same Fisher analysis, we estimate that the reconstructed $P_{\rm r}(k)$ improves on the prereconstruction $\alpha$ constraint by roughly 30\%. This is roughly twice the improvement we find using ten bispectrum triangle configurations. Both bispectrum and reconstruction perform better at lower redshift, as shown in Table~\ref{tab:improvement_redshift}. 
\begin{table}
\centering
\resizebox{1.0\columnwidth}{!}{
\begin{minipage}{1.0\columnwidth}
\centering
\caption{{\normalfont At all redshifts tested, reconstruction gives roughly twice the improvement in $\alpha$ of our ten bispectrum configurations.}}
\label{tab:improvement_redshift} 
\begin{tabular}{c c  c  c } 
\tableline \tableline
 Redshift & $n$  ($h^3/ {\rm Mpc}^{3}$) & Reconstruction & Bispectrum \\ \tableline
0.15 & $3.8 \times 10^{-4}$ & 38\% & 17\% \\  
0.55 & $3.8 \times 10^{-4}$ & 32\% & 14\%\\  
0.8 & $8.5 \times 10^{-4}$ & 26\%  & 10\% \\
 \tableline \tableline
\end{tabular}
\end{minipage}}
\end{table}
At all three redshifts, reconstruction still gives roughly twice the improvement of our ten bispectrum configurations. We note that additional configurations will likely narrow this gap \cite{2018arXiv180602853G}, and the comparison between bispectrum and reconstruction improvements does depend on the range of wave numbers used in the analysis. Decreasing the maximum wave number $k_{\rm max}$ from the value of $0.2 \, h/{\rm Mpc}$ used above, bispectrum information provides more improvement in $\sigma_\alpha$ while reconstruction provides less. The bispectrum starts to perform better than reconstruction below $k_{\rm max} \sim 0.15 \, h/{\rm Mpc}$; 
for example, at $k_{\rm max} = 0.14 \, h/{\rm Mpc}$ and $z=0.55$, ten bispectrum configurations provide a 20\% improvement, compared to 18\% from reconstruction.

{\sl Discussion.--} Measuring the bispectrum for only 190 triangles can improve constraints on the BAO scale by 14\%, corresponding to an increase in survey time of roughly 30\%. Our 14\% improvement with tailored triangles is comparable to the roughly 10\% improvement found by \cite{Pearson:2017wtw} using a more complete set of triangles, and the 6\% improvement of \cite{Slepian:2015qza}. More critically, our central finding is that the improvement depends not only on the \emph{number} of bispectrum measurements, but on the \emph{choice} of measurements. By selecting triangle configurations where interference effects amplify the BAO feature in the bispectrum, we obtain constraints with relatively few measurements, decreasing the number of mock catalogs needed to estimate the covariance matrix ${\bf C}$. Our method also opens a new avenue for numerically obtaining the cross covariance between $P$ and $B$, which is less easily treated with an analytic template than is the autocovariance. Additionally, our $(k_1, \delta, \theta)$ parametrization enables visualization of the bispectrum BAO feature in simple 1D plots (Fig.~\ref{fig:sim_destructive_constructive}). 

Reconstruction is a dynamical method that, applied to a single realization of the Universe, partially removes nonlinear effects using the full density field information---including information that is not captured even by higher-order statistics. Bispectrum measurements of BAOs are operationally independent from reconstruction, so agreement between the two methods will demonstrate robust measurements of the BAO scale. For example, the bispectrum provides an additional check for sources of error in reconstruction, such as those described in \cite{2018arXiv180804384S}: incorrect assumptions of bias, redshift-space distortions, or distance parameters. With better understanding of the covariance between the postreconstruction power spectrum and the bispectrum triangles most relevant for BAOs, it may be possible to combine pre- and postreconstruction measurements to further improve constraints on the BAO scale. Furthermore, the bispectrum can easily be extended to constrain BAOs even in modified gravity models. Once measured, bispectra can simply be compared to any modified gravity model; reconstruction, in contrast, must be modified according to each specific model and reapplied to the data before repeating power spectrum measurements to constrain BAOs. 

Our technique highlights phase effects; it thus may constrain sources of phase shifts in the power spectrum. One source of a phase shift is an isocurvature perturbation, where the oscillation is proportional to $\cos k\tilde{s}$ instead of $\sin k\tilde{s}$ \cite{1999PhRvD..59h3509H}. We expect that destructive configurations (as identified in the no-isocurvature model) may now show a BAO feature.
In future work, we will explore this phase shift as well as that induced by relativistic species such as neutrinos \cite{2015PhRvL.115i1301F, 2017JCAP...11..007B, Baumann:2018qnt}.

Future work will also discuss the dependence of BAO amplitude on the triangle parameters $\theta$ and $\delta$, explore whether additional triangle configurations offer any improvement in constraints, and study the independence of bispectrum information from that used in reconstruction. When combined with reconstruction, our result may represent a further improvement in BAO precision. Whether or not it does so on a statistical level, bispectrum measurements are operationally independent from reconstruction and therefore subject to different systematic effects. The bispectrum thus offers at the very minimum a cross-check that, added to those of \cite{2017MNRAS.470.2617A}, will be valuable for analysis of BAOs in future large-scale surveys.

\begin{acknowledgments}
{\sl Acknowledgments.--} 
M.~T. is supported in part by the World Premier International Research Center (WPI) Initiative, Ministry of Education, Culture, Sports, Science and Technology (MEXT), Japan, and the Japan Society for the Promotion of Science (JSPS) Grants-in-Aid for Scientific Research (KAKENHI) Grants No.~JP15H03654, No.~JP15H05887, No.~JP15H05893 and No.~JP15K21733. Z.~S. is supported by the National Aeronautics and Space Administration (NASA) through Einstein Postdoctoral Fellowship Award No. PF7-180167. T.~N. is supported in part by JSPS KAKENHI Grant No. 17K14273, and Japan Science and Technology Agency (JST) Core Research for Evolutional Science and Technology (CREST) Grant No. JPMJCR1414. H.~L.~C. was supported by the National Science Foundation (NSF) Graduate Research Fellowships Program (GRFP) under Grant No. DGE-1746045, an international travel allowance through Graduate Research Opportunities Worldwide, and an International Research Fellowship of JSPS [Postdoctoral Fellowships for Research in Japan (Strategic)]. The work of S.~H., K.~H.,  and  H.~L.~C.  at  Argonne National Laboratory was  supported  under  the  U.S.  Department of Energy (DOE) Contract No.~DE-AC02-06CH11357.  This research used resources of the National Energy Research Scientific Computing Center (NERSC), a DOE Office of Science User Facility supported by the Office of Science
of the U.S. DOE under Contract No.  DE-AC02-05CH11231.

\end{acknowledgments}

 \bibliography{refs}
\end{document}